\documentclass[twocolumn]{aastex63}
\usepackage{amsmath}
\usepackage{booktabs}

\newcommand{\correction}[1]{\textcolor{black}{#1}}

\submitjournal{ApJL}

\shorttitle{Detecting Molecules in Earth-like Exoplanets Orbiting White Dwarfs with JWST}

\shortauthors{Kaltenegger \& MacDonald et al.}

\begin{document}

\title{The White Dwarf Opportunity: Robust Detections of Molecules in Earth-like \\ Exoplanet Atmospheres with the James Webb Space Telescope}

\correspondingauthor{Lisa Kaltenegger 
\\ 
\\ 
$^*$Joint 1st authors:}
\email{lkaltenegger@astro.cornell.edu}
\email{rmacdonald@astro.cornell.edu}

\author[0000-0002-0436-1802]{Lisa Kaltenegger$^{*}$}
\affiliation{Carl Sagan Institute and Department of Astronomy, Cornell University, 122 Sciences Drive, Ithaca, NY 14853, USA}
\author[0000-0003-4816-3469]{Ryan J. MacDonald$^{*}$}
\affiliation{Carl Sagan Institute and Department of Astronomy, Cornell University, 122 Sciences Drive, Ithaca, NY 14853, USA}
\author[0000-0002-3868-2129]{Thea Kozakis}
\affiliation{Carl Sagan Institute and Department of Astronomy, Cornell University, 122 Sciences Drive, Ithaca, NY 14853, USA}
\author[0000-0002-8507-1304]{Nikole K. Lewis}
\affiliation{Carl Sagan Institute and Department of Astronomy, Cornell University, 122 Sciences Drive, Ithaca, NY 14853, USA}
\author[0000-0003-2008-1488]{Eric E. Mamajek}
\affiliation{Jet Propulsion Laboratory, California Institute of Technology, 4800 Oak Grove Drive, Pasadena, CA 91109, USA}
\affiliation{Department of Physics \& Astronomy, University of Rochester, Rochester, NY 14627, USA}
\author[0000-0002-7093-295X]{Jonathan C. McDowell}
\affiliation{Harvard-Smithsonian Center for Astrophysics, 60 Garden St, Cambridge, MA 02138, USA}
\author[0000-0001-7246-5438]{Andrew Vanderburg}
\affiliation{Department of Astronomy, The University of Texas at Austin, Austin, TX 78712, USA}
\affiliation{NASA Sagan Fellow}

\begin{abstract}

\noindent 
The near-term search for life beyond the solar system currently focuses on transiting planets orbiting small M dwarfs, and the challenges of detecting signs of life in their atmospheres. However, planets orbiting white dwarfs (WDs) would provide a unique opportunity to characterize rocky worlds. The discovery of the first transiting giant planet orbiting a white dwarf, WD 1856+534b, showed that planetary-mass objects can survive close-in orbits around WDs. The large radius ratio between WD planets and their host renders them exceptional targets for transmission spectroscopy. Here, we explore the molecular detectability and atmospheric characterization potential for a notional Earth-like planet, evolving in the habitable zone of WD 1856+534, with the James Webb Space Telescope (JWST). We establish that the atmospheric composition of such Earth-like planets orbiting WDs can be precisely retrieved with JWST. We demonstrate that robust $> 5\,\sigma$ detections of H$_2$O and CO$_2$ can be achieved in a 5-transit reconnaissance program, while the biosignatures O$_3$ + CH$_4$, and O$_3$ + N$_2$O can be detected to $> 4\,\sigma$ in as few as 25 transits. N$_2$ and O$_2$ can be detected to $> 5\,\sigma$ within 100 transits. Given the short transit duration of WD habitable zone planets ($\sim$ 2\,minutes for WD 1856+534), conclusive molecular detections can be achieved in a small or medium JWST transmission spectroscopy program. Rocky planets in the WD habitable zone therefore represent a promising opportunity to characterize terrestrial planet atmospheres and explore the possibility of a second genesis on these worlds.

\end{abstract}


\section{Introduction}
\label{sec:intro}

The discovery of a planetary-mass object orbiting WD~1856+534 \citep{Vanderburg2020} shows that such objects can survive close-in ($\lesssim$ 0.02\,AU) orbits around white dwarfs (WDs). This complements the recent detection of a gaseous debris disk from a giant planet orbiting WD~J0914+1914 \citep{gans19}. It is well established that 25-50\% of WDs display spectral signatures from recent metal pollution \citep[e.g.][]{Koester2014}, a sign of rocky bodies scattered toward the host resulting in orbiting planetary debris (e.g.\ \citealt{Jura2014,Vanderburg2015,Bonsor2020}). These discoveries indicate rocky bodies exist in WD systems, motivating searches for terrestrial planets around WDs \citep[e.g.][]{Fulton2014,vanSluijs2018}.

The origin and survival of close-in planets orbiting WDs has seen active theoretical study. Once a main sequence star evolves into a WD, stable planetary systems can undergo violent dynamical instabilities \citep{Debes2002}, exciting planets into high eccentricity, low pericentre, orbits \citep{Veras2015}. These orbits can rapidly circularize due to tidal dissipation, leading in some circumstances to the survival of planets in close-in orbits \citep{Veras2019a}.

The possibility of habitable planets orbiting WDs has been discussed by several authors (e.g. \citealt{mccr71,Agol11,foss12,loeb13,Kozakis2018,Kozakis2020}). Though such a planet has yet to be discovered, observations with NASA’s K2 mission \citep{Howell2014} have constrained the occurrence rate of Earth-sized habitable zone (HZ) planets around nearby WDs to be $<$ 28\% \citep{vanSluijs2018}. Planets orbiting WDs experience relatively stable environments for billions of years after initial cooling. An average WD spends several billion years cooling from 6,000\,K to 4,000\,K \citep{berg01}, providing planets in the WD HZ with a habitable timescale nearly twice that expected for Earth \citep{Kozakis2018}.

Spectroscopic observations of giant planets, mini-Neptunes, and super-Earths around main sequence stars have yielded detections of molecular, atomic, and ionic species in dozens of planets \citep[e.g.][]{Charbonneau2002,Deming2013,Hoeijmakers2018,Benneke2019a}. Transiting planets orbiting smaller stars are generally easier to characterize, due to their increased planet-to-star size ratio. Consequently, the smallest worlds characterized to date, including K2-18b \citep{Benneke2019b,Tsiaras2019} and LHS 3884b \citep{Kreidberg2019}, reside around small stars.

Proposals to extend the `small star opportunity' to HZ terrestrial planets have focused on small M-type stars. Discoveries such as Proxima b \citep{Anglada2016} and the TRAPPIST-1 system \citep{Gillon2017} have resulted in numerous studies of their atmospheric characterization potential \citep[e.g.][]{Barstow2016,Morley2017,Tremblay2020,Lin2020}. The James Webb Space Telescope (JWST) offers a near-term avenue to access these atmospheres. However, while some molecular detections, such as CO$_2$ and H$_2$O, can be achieved for HZ planets in $\sim$ 10 transits for close-by M dwarf planets \citep{Krissansen-Totton2018,Lustig-Yaeger2019}, the detection of biosignatures, such as O$_3$ or O$_2$ combined with a reducing gas like CH$_4$ \citep[see][for a recent review]{kalt17} will be at the limit of JWST's capability over its lifetime \citep{Lustig-Yaeger2019}.

White dwarfs, similar in size to Earth, offer even better contrast ratios than M dwarfs, rendering rocky planets around WDs promising targets for atmospheric characterization \citep{Agol11,loeb13,Kozakis2020}. While small planets orbiting WDs could be second-generation planets and provide different conditions than on Earth \citep[see e.g. discussion in ][]{Kozakis2018}, characterizing WD planets would answer intriguing questions on lifespans of biota (e.g.\ \citealt{Sagan1993,omal2013,kalt17}) or a second “genesis” after a star’s death. However, a detailed analysis examining the detectability of specific molecular species, including biosignatures, in the atmospheres of HZ planets around WDs has yet to be undertaken. 

In this letter, we demonstrate the atmospheric characterization potential with JWST for a notional Earth-like planet which evolved around a WD. In what follows, we generate a model transmission spectrum, produce synthetic JWST observations, and conduct an extensive atmospheric retrieval analysis.

\section{The Atmospheres of Earth-like planets orbiting White dwarfs} \label{sec:WD_planet_atmosphere}

\begin{figure*}[ht!]
     \centering
     \includegraphics[width=\textwidth]{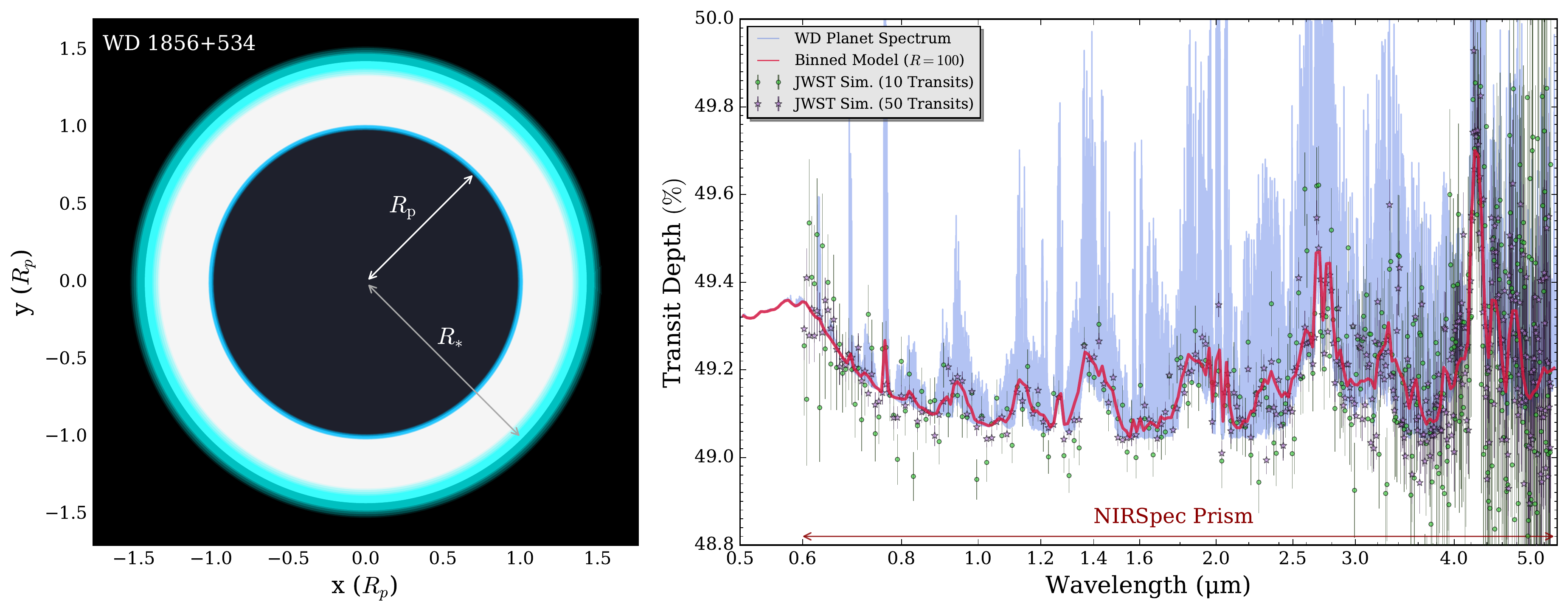}
     \caption{Transmission spectroscopy of an Earth-like planet orbiting a white dwarf. Left: transit geometry for an Earth-radius exoplanet in the WD 1856+534 system (observer's perspective). The WD (white disk), planet (grey disk), and planetary atmosphere (blue annulus) are to scale. This geometry has a fractional flux decrement during transit, or transit depth, of $\sim 50$\%. Right: corresponding model transmission spectrum and synthetic JWST observations. The high resolution model ($R =$ 50,000, blue) is binned to a low resolution ($R = 100$, red) for comparison with simulated JWST NIRSpec Prism observations.}
     \label{fig:geometry+spectrum}
\end{figure*}

Consider a notional planet with the radius and mass of Earth residing in the HZ of WD~1856+534. Given the incident flux and orbital parameters of WD~1856+534b ($S = 0.191\,S_{\earth}$, $a/R_{*} = 326.9$, $P = 1.41$\,day, \citet{Vanderburg2020}), scaling to the irradiation of the modern Earth ($S = S_{\earth}$) yields $a/R_{*} = 142.86$ and $P = 9.8$\,hr. Residing at 2.9 Roche radii, such a planet would experience strong tidal forces -- with long-term survival largely contingent on the planetary viscosity \citep[see][]{Veras2019b}. Our nominal Earth-like planet, orbiting a WD with $R_{*} = 0.0131\,R_{\sun}$, has $R_{\rm p}/R_{*} = 0.6995$, a transit depth of $(R_{\rm p}/R_{*})^2 \approx 49\%$, transit probability of $\approx 1.2\%$, and a transit duration of 2.2\,min. The geometry of our system is shown in Figure~\ref{fig:geometry+spectrum} (left panel). 

\subsection{Atmospheric Models} \label{subsec:atmospheric_models}

We model the atmosphere of our Earth-like exoplanet, which evolved around a WD, using EXO-Prime \citep{Kaltenegger2010} -- a 1D radiative-convective terrestrial atmosphere code. Given an incident stellar / host spectrum and planetary outgassing rates, EXO-Prime couples 1D climate and photochemistry models to compute the vertical temperature structure and atmospheric mixing ratio profiles. The application of EXO-Prime to Earth-like planets evolving around WD hosts from $T_{\rm eff} = 6000 - 4000$\,K is extensively described in \citet{Kozakis2018} and \citet{Kozakis2020}. Here, we define `Earth-like' to refer to an Earth radius and Earth mass planet with similar outgassing rates to the modern Earth.

The irradiation environment around a WD changes the atmospheric composition of an Earth-like planet compared to the modern Earth \citep{Kozakis2018}. Given the temperature of WD ~1856+534 ($T_{\rm eff} = 4710\,$K, \citet{Vanderburg2020}), we use the 5000\,K WD HZ terrestrial planet model from \citet{Kozakis2020} for our analysis. We note that WD hosts display largely featureless spectra\footnote{For our input stellar spectrum, see the \href{https://www.doi.org/10.5281/zenodo.3960468}{supplementary material}.} \citep{Saumon2014}, similar to black bodies, below 5000\,K \citep{Kepler2016}. The atmosphere of a rocky planet receiving $S = S_{\earth}$ around a 5000\,K WD host, with Earth-like outgassing rates, shows higher levels of CH$_4$, lower O$_3$, and similar H$_2$O concentrations compared to Earth, along with a temperature inversion. We note that the surface UV environment of a WD planet is time dependent, impacting the atmospheric composition \citep[see][]{Kozakis2018}. We tested our model sensitivity to UV levels by artificially increasing the flux in our Ly-$\alpha$ bin (1200 - 1300 {\AA}) by 1000$\times$. Such an unphysical increase only slightly lowers the CH$_4$, H$_2$O, O$_3$, and N$_2$O abundances in the upper atmosphere ($<$ a factor of 3), so our results are robust to higher UV environments expected around younger WDs. A detailed account of our climate-photochemistry model is provided in \citet{Kozakis2018}.

\subsection{Transmission Spectrum} \label{subsec:transmission_spectrum}

Our model transmission spectrum is computed with the POSEIDON radiative transfer code \citep{MacDonald2017}. POSEIDON has been widely applied to the modeling and interpretation of giant planet atmospheres, ranging from hot Jupiters to exo-Neptunes \citep[e.g.][]{Sedaghati2017,Kilpatrick2018,MacDonald2019}. For the present study, we extended the opacity database in POSEIDON to include all molecular species with prominent absorption features expected in the atmospheres of Earth-like planets orbiting WDs \citep{Kozakis2020}. We pre-computed cross sections for O$_2$, O$_3$, H$_2$O, CH$_4$, N$_2$O, CO$_2$, CO, and CH$_3$Cl from HITRAN2016 line lists \citep{Gordon2017} using the HITRAN Application Programming Interface (HAPI) \citep{Kochanov2016}. Our cross section grid covers temperatures from $100-400\,$K (20\,K spacing), pressures from $10^{-6} - 10\,$bar (1 dex spacing), and wavelengths from $0.4-50\,\micron$ (0.01\,cm$^{-1}$ resolution). All spectral lines are modeled as air-broadened Voigt profiles, with the line wings calculated to either 500 half-width half maxima or 30 cm$^{-1}$ from the line core (whichever deviation is smaller). Given the importance of optical wavelength O$_3$ opacity in terrestrial planet transmission spectra \citep[e.g.][]{kalt17, Meadows2018,Kozakis2020}, not currently included in HITRAN, we employ the temperature-dependent O$_3$ cross sections from \citet{Serdyuchenko2014}. The latest HITRAN Collision-induced Absorption (CIA) data, covering N$_2$
-N$_2$, O$_2$-O$_2$, O$_2$-N$_2$, N$_2$-H$_2$O, O$_2$-CO$_2$, CO$_2$-CO$_2$, and CO$_2$-CH$_4$ is also included \citep{Karman2019}.

A high-resolution transmission spectrum is derived from radiative transfer through our model Earth-like atmosphere. The equation of radiative transfer is solved by integrating the wavelength-dependent extinction coefficient along the line of sight for successive atmospheric annuli. The mixing ratio profiles and temperature structure described in Section~\ref{subsec:atmospheric_models} were interpolated onto an altitude grid with 10 layers per pressure decade, uniformly spaced in log-pressure, from the surface to $10^{-7}\,$bar. For layers above 0.1\,mbar, the atmosphere was assumed to posses the same temperature and composition as present at 0.1\,mbar\footnote{The resulting temperature and mixing ratio profiles are shown in the \href{https://www.doi.org/10.5281/zenodo.3960468}{supplementary material}.}. The resultant high resolution spectrum matches the spectra modeled by \citet{Kozakis2020} -- with the addition of CIA, which has been added to the transmission spectra presented here.

We modify our radiative transfer prescription to correct for the effect of refraction. Light rays probing sufficiently dense, deep regions of an atmosphere do not contribute to transmission spectra due to refraction preventing rays reaching a distant observer \citep[e.g.][]{Betremieux2014,Robinson2017b}. For a given planet and atmosphere, this effect is most prominent where the angular size of a star from the perspective of its planet is small. For example, a remote observation of the Earth-Sun system would yield a transmission spectrum that can only be probed down to altitudes of about 12.7\,km \citep[e.g.][]{Betremieux2014}. In the case of our WD planet ($R_{*}/a < 10^{-2}$), we employ equation 14 from \citet{Robinson2017b} to estimate a refractive surface occurs at 0.523\,bar (5.2\,km altitude). We simulate the effect of refraction by neglecting the contribution of impact parameters below the refractive surface \citep[e.g.][]{Betremieux2014,Robinson2017b,MacdonaldE2019}. Following this procedure, we computed a high-resolution ($R =$ 50,000) transmission spectrum from $0.4-5.4\,\micron$, shown in Figure~\ref{fig:geometry+spectrum} (right panel).

\section{Observing Earth-like Planets Orbiting White Dwarfs with JWST} \label{sec:observations}

Observational spectroscopy of transiting planets orbiting WDs presents unique opportunities and challenges compared to main sequence stars. First, there are the intrinsic challenges of the presently unknown occurrence rate of planets in the WD HZ \citep{vanSluijs2018} and the geometric probability of observable transits ($\approx 1\%$). Second, there are technical challenges of fainter hosts (J $>$ 13) and transit durations on minute timescales. On the other hand, a distinct opportunity is intrinsically strong atmospheric molecular features due to the high $R_{\rm{p}}/R_{*}$ ($\gtrsim 50\%$). Should such a transiting planet be detected in future, here we demonstrate that the predicted strength of atmospheric signatures overcomes the aforementioned technical issues, resulting in highly favorable transmission spectroscopy with JWST.

\subsection{Observational Strategy} \label{subsec:observing_strategy}

Imagine now observing an Earth-like planet transiting WD 1856+534 with JWST. With a 2.2\,min transit duration, and a twice-transit out of transit baseline, a minimum exposure time of 6.6\,min per transit results. However, JWST time constrained observations\footnote{Observations in which the commencement window acceptable margin is $<$ 1 hr, see \href{https://jwst-docs.stsci.edu/jwst-opportunities-and-policies/jwst-call-for-proposals-for-cycle-1/jwst-cycle-1-observation-types-and-restrictions/time-constrained-observations}{JWST Cycle 1 documentation}.} incur a 1 hr penalty per visit to ensure a sufficient baseline to characterize instrument systematics. We therefore conservatively allocate 1.5\,hr observing time per transit. Further allowing a 40\% overhead\footnote{Based on approved program \href{http://www.stsci.edu/jwst/observing-programs/program-information?id=1331}{\#1331} (PI: Nikole Lewis).} above the science time, we estimate 12 transits can fit in a small program ($<$ 25\,hr), 35 in a medium program ($<$ 75\,hr) and 35 + in a large program ($>$ 75\,hr). In what follows, we consider the atmospheric characterization potential of representative small, medium, and large JWST programs.

\begin{figure*}[ht!]
     \centering
     \vspace{-0.4cm}
     \includegraphics[width=0.96\textwidth, trim={0.2cm 0.2cm 0.2cm -0.4cm}]{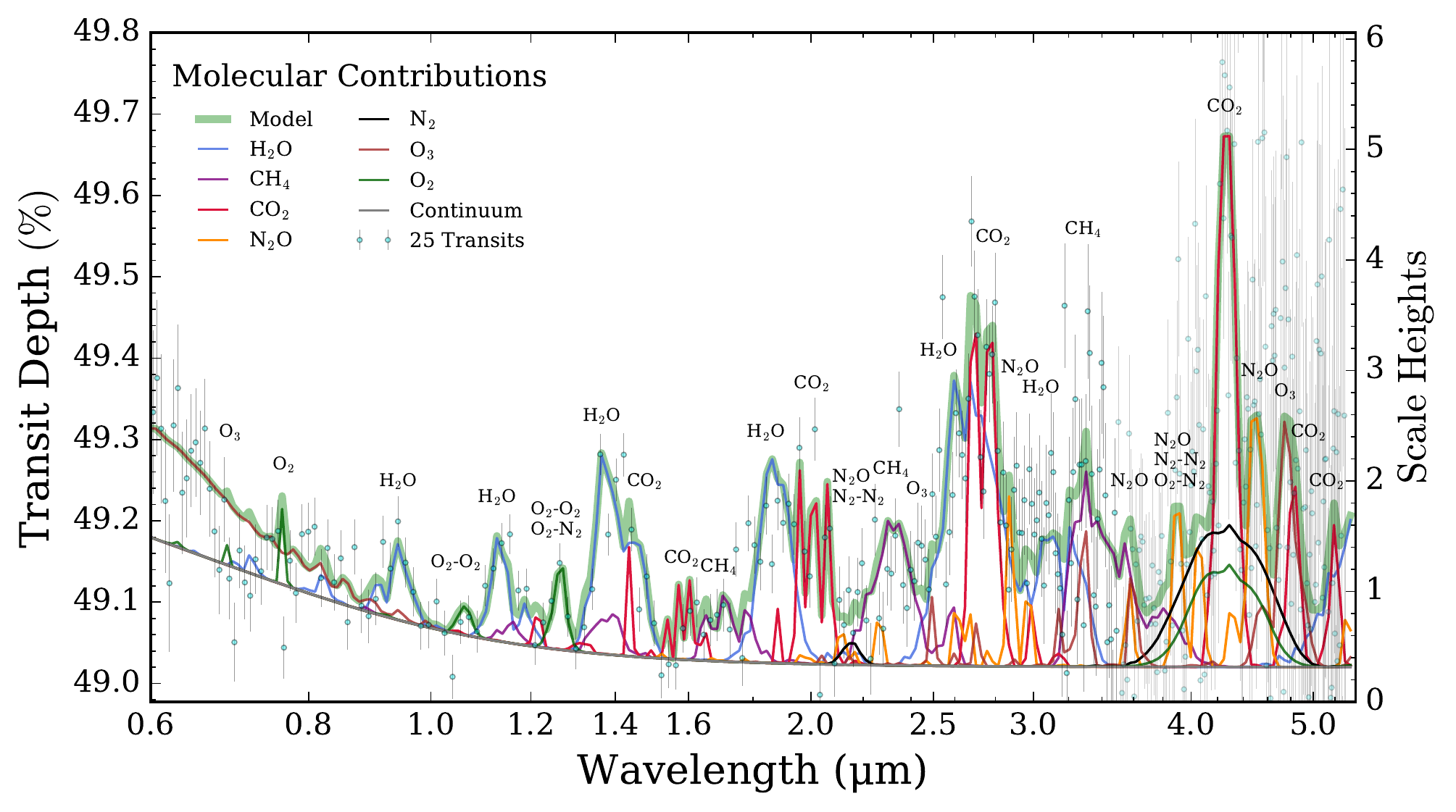}
     \caption{Molecular contributions to the transmission spectrum of an Earth-like exoplanet orbiting WD 1856+534. We display the best-fitting transmission spectrum retrieved from a simulated 25 transit JWST program (green shading). The opacity contributions of each retrieved molecule (colored curves) are shown, relative to the spectral continuum due to Rayleigh scattering and refraction (gray curve). Prominent absorption features are labeled. Collision-induced absorption (CIA) pairs featuring O$_2$ (O$_2$-O$_2$ and O$_2$-N$_2$) are depicted alongside the O$_2$ contribution. N$_2$-N$_2$ CIA provides a contribution around $4\,\micron$. All contributions are plotted at $R=100$ for ease of comparison with the simulated JWST observations (error bars). The number of equivalent scale heights ($H \approx 8.8$\,km) above the surface, for which an opaque atmosphere at a given wavelength would produce the same transit depth, is shown on the right. For such observations, H$_2$O, CO$_2$, CH$_4$, and O$_3$ are confidently detected to $> 5 \sigma$ confidence, N$_2$O to $> 4 \sigma$, and O$_2$ to $> 2 \sigma$.}
     \label{fig:molecular_features}
\end{figure*}

\subsection{Simulated JWST Observations} \label{subsec:pandexo_data}

We simulate JWST observations with the NIRSpec Prism using PandExo \citep{Batalha2017}. The NIRSpec Prism covers a wide spectral range ($0.6 - 5.3\,\micron$) in a single transit, yielding high information content for terrestrial exoplanet atmospheres \citep{Batalha2018}. We focus on the Prism due to its coverage of a wide array of spectral features, including O$_2$, O$_3$, H$_2$O, CO$_2$, CH$_4$, and N$_2$O (see Figure~\ref{fig:molecular_features}). Other instrument modes, such as NIRISS SOSS, may provide complementary information in narrower spectral regions. We generated synthetic JWST observations for 1, 2, 3, 5, 10, 25, 50, and 100 transits using the NIRSpec Prism's 512 subarray for the model transmission spectrum from Section~\ref{subsec:transmission_spectrum}. Example noise instances for the 10 and 50 transit cases are shown in Figure~\ref{fig:geometry+spectrum} (right panel). For the host WD spectrum, we assume a 4780\,K black body, normalized to J = 15.677 \citep{Cutri2003}. For a total integration time (inside and outside transit) of 1.5\,hr, PandExo's optimizer determined a $\sim$ 38\,s exposure time and 166 groups per integration. We generated simulated observations at their native resolution ($R \sim 30-300$) to avoid information loss from binning \citep[see][]{Benneke2013,Tremblay2020}.

The spectral uncertainties generated by PandExo can be used in two ways. One approach produces simulated observations for a specific (Gaussian) noise instance (e.g. Figure~\ref{fig:geometry+spectrum}). However, as noted by \citet{Feng2018}, results derived whenceforth are specific to this noise instance. They showed that a dataset with the same uncertainties, but centered on the model, results in derived probability distributions representative of the average over many noise instances (via the central limit theorem). We follow this second approach in reporting our predicted detection significances and abundance constraints, ensuring our results are unbiased by individual noise realizations. Nevertheless, we verified that consistent results occur using data with Gaussian scatter. We now subject our simulated JWST datasets to a series of atmospheric retrievals, assessing the ability to recover atmospheric information from such observations.

\section{Characterization of White Dwarf Planet Atmospheres with JWST} \label{sec:retrievals}

Here we present the first detailed study of the atmospheric characterization potential for Earth-like planets orbiting WDs with JWST. We employ the technique of \emph{atmospheric retrieval} to explore the range of models consistent with synthetic JWST observations. In what follows, we first outline our retrieval process, before presenting predicted detection significances and mixing ratio constraints for the most abundant molecules in our Earth-like atmosphere around a WD.

\subsection{Atmospheric Retrieval Procedure} \label{subsec:retrieval_methods}

Atmospheric retrieval refers to the inversion of atmospheric properties from a planetary spectrum. Retrieval codes couple a parametric atmospheric model and radiative transfer solver with a statistical sampling algorithm. Retrieval techniques are commonly applied in Solar System remote sounding \citep[e.g.][]{Irwin2018} and to spectroscopy of giant exoplanets \citep[see][for a review]{Madhusudhan2018}. Here, we employ atmospheric retrievals to rigorously conduct Bayesian parameter estimation and nested model comparisons. 

Our retrievals are computed using the retrieval module of POSEIDON \citep{MacDonald2017}. This utilizes a parametric version of the transmission spectrum model described in Section~\ref{subsec:transmission_spectrum}, coupled with the nested sampling algorithm MultiNest \citep{Feroz2008,Feroz2009,Feroz2013}. The atmosphere is divided into 81 layers uniformly spaced in log-pressure from $10^{-7} - 10\,$bar. The mixing ratios of O$_2$, O$_3$, H$_2$O, CO$_2$, CH$_4$, N$_2$O, CO, and CH$_3$Cl, assumed uniform in altitude, are ascribed as free parameters, with the remaining gas composed of N$_2$. Each mixing ratio has a log-uniform prior from $10^{-12} - 1$, with mixing ratio sums exceeding unity rejected. The temperature structure is parametrized by the six-parameter function from \citet{Madhusudhan2009}, with priors as in \citet{MacDonald2019} but for a surface (1\,bar) temperature parameter (uniform prior from $100 - 400\,$K). Two parameters are also assigned for the planetary radius at 1\,bar (uniform prior from $0.8 - 1.2\,R_{\earth}$) and the pressure of the refractive surface (log-uniform prior from $10^{-7} - 10\,$bar). This yields a total of 16 free parameters. Radiative transfer is evaluated at $R = 2000$ from $0.5 - 5.4\,\micron$, with each model spectrum convolved to the resolving power of the NIRSpec Prism and integrated over its sensitivity function for comparison with each data point. The parameter space exploration is conducted via the PyMultiNest package \citep{Buchner2014} with 2,000 live points.

We conducted eight atmospheric retrievals for each synthetic JWST dataset, for a total of 64 retrievals. For each number of transits, one `full' retrieval was computed for parameter estimation, along with seven retrievals each excluding one molecule. These nested retrievals yield prediction detection significances for N$_2$, O$_2$, O$_3$, H$_2$O, CO$_2$, CH$_4$, and N$_2$O (CO and CH$_3$Cl are unconstrained in all cases) via Bayesian model comparisons \citep[see][]{Trotta2017}. A typical retrieval involved the computation of 1-5 million transmission spectra. We show an example of the best-fitting spectrum from the 25 transit case in Figure~\ref{fig:molecular_features}, demonstrating the ability to correctly infer the principal molecular species present in our model atmosphere. We now proceed to present the results from our series of atmospheric retrievals.

\begin{figure*}[ht!]
     \centering
     \includegraphics[width=0.985\textwidth, trim={0.0cm 0.2cm 0.0cm 0.0cm}]{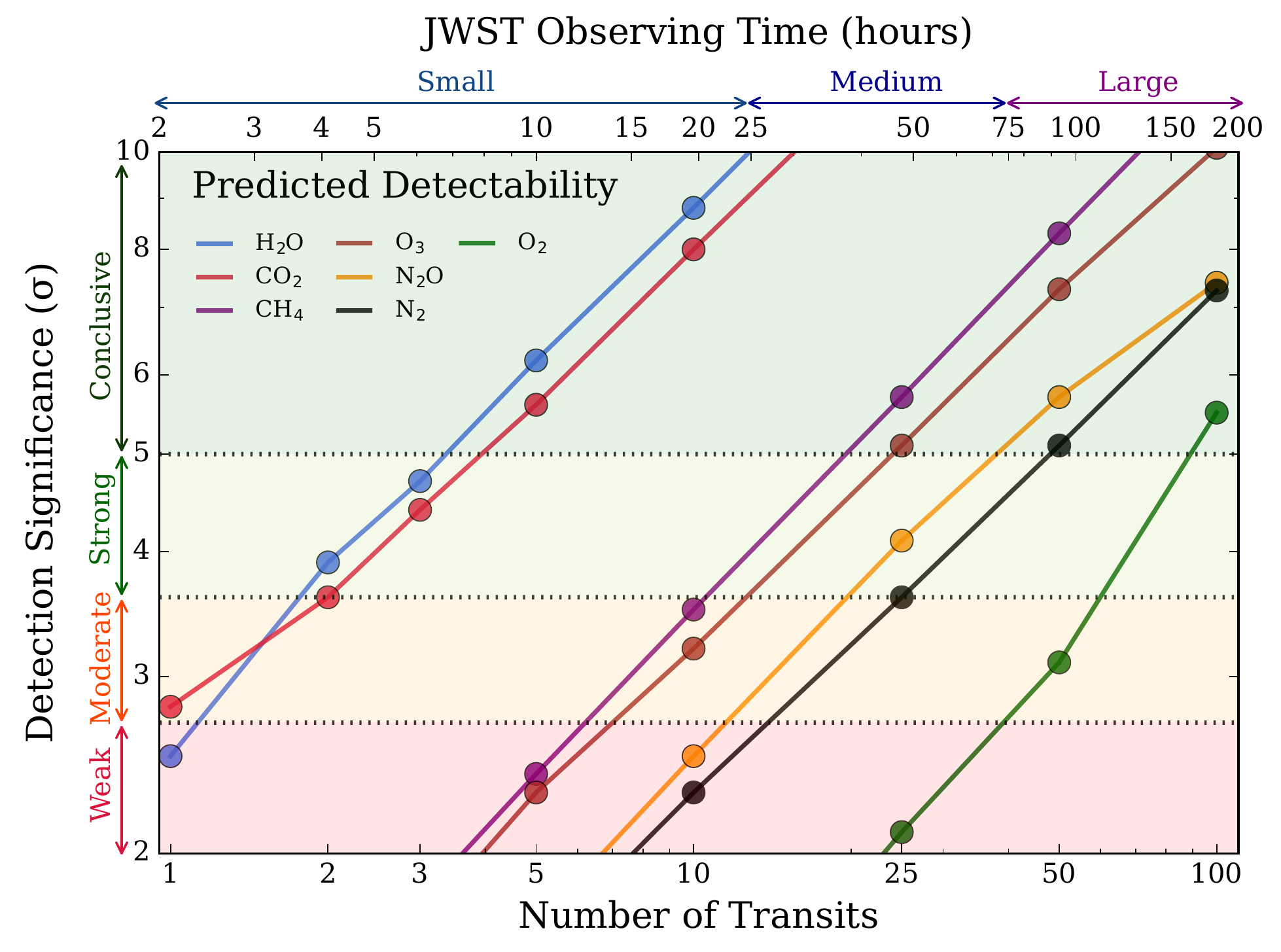}
     \caption{Predicted detection significances of atmospheric molecular species for an Earth-like planet orbiting WD 1856+534. The predictions are given as a function of the number of transits (2.2\,min duration) and corresponding JWST observing time (including time constrained observation charges and overheads, see section~\ref{subsec:observing_strategy}). The detection significances are sorted into `weak', `moderate', `strong', and `conclusive' detections (shaded regions), according to an adaptation of the Jeffreys' scale for Bayesian model comparisons \citep[e.g.][]{Trotta2017}. The boundaries (dotted lines) occur at 2.7$\sigma$, 3.6$\sigma$, and 5.0$\sigma$, respectively. Within a 10 transit small program ($\sim$ 20\,hr), H$_2$O and CO$_2$ can be conclusively detected, with CH$_4$ and O$_3$ detected to $> 3\sigma$. A 25 transit medium program ($\sim$ 50\,hr) can conclusively detect CH$_4$ and O$_3$, while additionally detecting N$_2$O to $> 4\sigma$. A further increase to a 100 transit large program ($\sim$ 200\,hr) yields conclusive detections of N$_2$ and O$_2$.}
     \label{fig:detections_vs_transits}
\end{figure*}

\begin{figure*}[ht!]
     \centering
     \includegraphics[width=\textwidth, trim={0.0cm 0.0cm 0.0cm 0.0cm}]{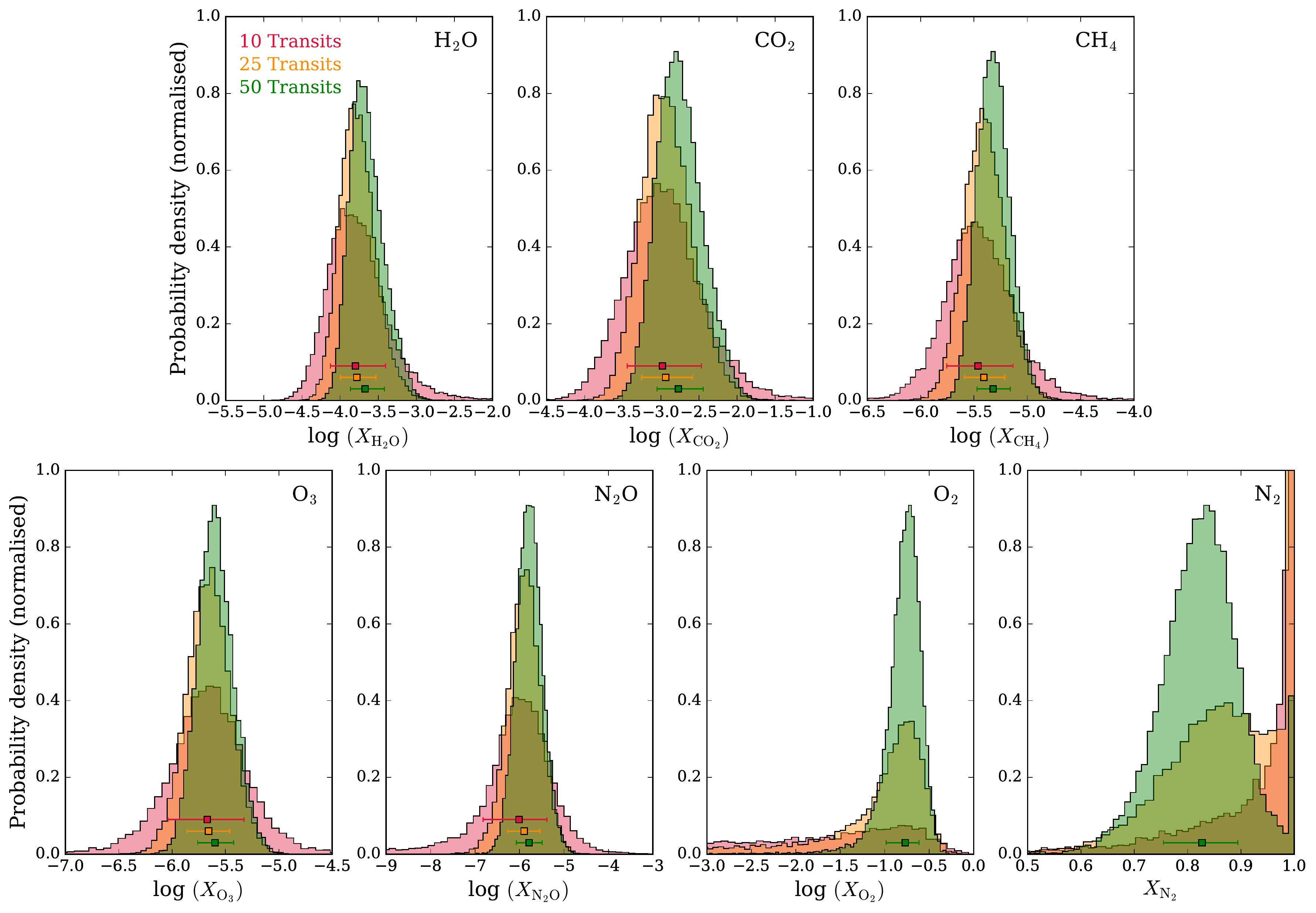}
     \caption{Predicted atmospheric composition constraints for an Earth-like planet orbiting WD 1856+534 with JWST. Marginalized posterior probability distributions for the abundances of each detectable molecule are shown, considering 10 transit (red), 25 transit (orange), and 50 transit (green) programs. The error bars give the median retrieved abundances and $\pm 1\sigma$ constraints. H$_2$O, CO$_2$, CH$_4$, O$_3$, and N$_2$O can be constrained to $<$ 1 dex (an order of magnitude) with 10 transits, $<$ 0.4 dex with 25 transits, and $<$ 0.3 dex (a factor of 2) within 50 transits. Placing a lower bound on the O$_2$ abundance is not possible for the 10 and 25 transit cases, hence the preferred solution is an N$_2$ dominated atmosphere ($X_{\rm{N_2}} \rightarrow 100$\%). Bounded constraints on the O$_2$ and N$_2$ abundances require at least 50 transits, after which they can be measured with an uncertainty of $\lesssim 7$\%.}
     \label{fig:abundance_posteriors}
\end{figure*}

\subsection{The Detectability of Atmospheric Molecules in Terrestrial Planets Orbiting White Dwarfs} \label{subsec:detection_significances}

Many atmospheric molecules, including prospective biosignatures, can be readily detected in Earth-like planets orbiting WDs with JWST. Our predicted detection significances are shown in Figure~\ref{fig:detections_vs_transits}, categorized according to an adaptation of the Jeffreys' scale for Bayesian model comparisons \citep[e.g.][]{Trotta2017}. The easiest molecules to detect are H$_2$O and CO$_2$, owing to their multiple strong absorption features throughout the infrared (see Figure~\ref{fig:molecular_features}). A single transit is sufficient to provide evidence for H$_2$O and CO$_2$ at $> 2\sigma$. A 5 transit reconnaissance program can yield $> 5\sigma$ detections of H$_2$O and CO$_2$, providing guidance for subsequent larger programs. Detecting other species becomes possible with only a few additional transits. CH$_4$ and O$_3$ are detectable at $> 3\sigma$ in 10 transits, accessible within a small JWST program. The strong detectability of O$_3$ is enabled by the prominent Chappuis and Wulf bands at visible wavelengths \citep[see][]{Serdyuchenko2014}, with a lesser contribution from the 4.8\,$\micron$ feature. N$_2$O becomes detectable at $> 4\sigma$ in 25 transits, owing to its features around 2.9, 3.9, and 4.5\,$\micron$. Detecting N$_2$ and O$_2$ is more challenging, with $> 5\sigma$ detections requiring 50 transits for N$_2$ and 100 transits for O$_2$, respectively.  

It is therefore possible to detect multiple \emph{combinations} of biosignatures, namely CH$_4$ in combination with O$_3$ and CH$_4$ in combination with N$_2$O \citep[see][]{kalt17} within the remit of a medium JWST program.

The absorption features due to CIA are crucial to correctly identify the dominant atmospheric constituents. The detectability of N$_2$ stems from the relative strengths of O$_2$ and N$_2$ collisional pairs, as an O$_2$ dominated atmosphere would produce strong O$_2$-O$_2$ features at 1.06 and 1.25\,$\micron$, inconsistent with the simulated data. Without these CIA features, distinguishing between O$_2$ and N$_2$ is challenging due to their similar molecular weights \citep{Benneke2013}. A secondary contribution to N$_2$ detectability is the broad N$_2$-N$_2$ feature around 4.3\,$\micron$ \citep[e.g.][]{Schwieterman2015}. The detectability of O$_2$, even at the low resolution of the NIRSpec Prism, arises from the combination of the A band (0.76\,$\micron$), O$_2$-O$_2$ (1.06, 1.25\,$\micron$), and O$_2$-N$_2$ (1.25, 4.3\,$\micron$) features.

\subsection{Constraining the Molecular Composition of Earth-like Planets Orbiting White Dwarfs} \label{subsec:abundance_constraints}

Beyond detecting molecular species, transiting WD planets would provide the opportunity to precisely measure molecular abundances in Earth-like atmospheres. Our predicted abundance constraints\footnote{Full posteriors are available in the \href{https://www.doi.org/10.5281/zenodo.3960468}{supplementary material}.} for representative small (10 transit), medium (25 transit), and large (50 transit) JWST programs are shown in Figure~\ref{fig:abundance_posteriors}.

The abundances of spectrally prominent trace gases can be constrained with only a handful of transits. A single transit provides relatively loose constraints on H$_2$O and CO$_2$ abundances, spanning 2 orders of magnitude to $1\sigma$. Constraints improve considerably with 5 transits, with H$_2$O, CO$_2$, CH$_4$, and O$_3$ measurable to within an order of magnitude. Within the remit of a small JWST program (10 transits), N$_2$O can also be constrained within an order of magnitude. A medium program (25 transits) results in abundance constraints approaching a factor of 2 in precision (0.3 dex). Finally, a large 100 transit program can measure most abundances to $\lesssim$ 0.15 dex (40\% precision). CO and CH$_3$Cl are unconstrained in all cases, due to their low abundances and negligible contributions to the true model spectrum. The predicted values for precise trace gas abundance constraints ($\leq$ 1 dex), as a function of the number of transits, are as follows:

\begin{itemize}
    \item H$_2$O: $N_{\rm trans} =$ 5 (0.6 dex), 10 (0.4 dex), 25 (0.24 dex), 50 (0.22 dex), 100 (0.12 dex).
    \item CO$_2$: $N_{\rm trans} =$ 5 (0.8 dex), 10 (0.5 dex), 25 (0.34 dex), 50 (0.30 dex), 100 (0.16 dex).
    \item CH$_4$: $N_{\rm trans} =$ 5 (0.6 dex), 10 (0.3 dex), 25 (0.19 dex), 50 (0.16 dex), 100 (0.09 dex).
    \item O$_3$: $N_{\rm trans} =$ 5 (1.0 dex), 10 (0.4 dex), 25 (0.20 dex), 50 (0.17 dex), 100 (0.11 dex).
    \item N$_2$O: $N_{\rm trans} =$ 10 (0.8 dex), 25 (0.36 dex), 50 (0.29 dex), 100 (0.19 dex).
\end{itemize}

Constraining the composition of the background gases, N$_2$ and O$_2$, is comparatively challenging. An upper bound on the O$_2$ abundance is possible even with 10 transits, due to the opacity contributions of O$_2$-O$_2$ CIA. However, the lack of a lower bound on the O$_2$ abundance for $\lesssim$ 25 transits results in the preferred solution being an N$_2$ dominated atmosphere. Once sufficiently precise observations are obtained, the abundances of N$_2$ and O$_2$ can be correctly inferred and precisely constrained. With a 50 transit program, the fractions of O$_2$ and N$_2$ in the atmosphere, $X_{\rm O_2}$ and $X_{\rm N_2}$, can be measured with uncertainties of $\sim$ 7\%. This improves to $\sim$ 4\% for a 100 transit program. It would therefore be eminently possible to precisely measure the molecular composition of Earth-like planets orbiting WDs.

\section{Summary \& Discussion} \label{sec:Discussion}

Terrestrial planets orbiting white dwarfs would offer an unprecedented opportunity to undertake detailed characterization of rocky exoplanets in the imminent future. Prompted by the discovery of a planetary-mass object orbiting WD 1856+534, we explored the molecular detectability and atmospheric characterization potential for a notional Earth-like planet that evolved in the same system. We have demonstrated that robust atmospheric detections, and precise abundance constraints, for such a planet can be readily obtained with JWST.

Our main results are as follows:

\newpage

\begin{enumerate}
    \item Robust ($> 5\,\sigma$) detections of H$_2$O and CO$_2$ can be obtained from JWST NIRSpec Prism transmission spectra of an Earth-like planet within a 5 transit JWST reconnaissance program.
    \item Biosignatures, including CH$_4$ + O$_3$ and O$_3$ + N$_2$O, can be detected to $> 4\,\sigma$ in 25 transits.
    \item The abundances of these trace molecules can be measured  to $\lesssim$ a factor of 2 within 25 transits.
    \item Both O$_2$ and N$_2$ can be detected to $> 5\,\sigma$ within 100 transits. Their detectability with the NIRSpec Prism is crucially enabled by absorption features due to collisional pairs (i.e. O$_2$-O$_2$ and N$_2$-N$_2$).
\end{enumerate}

We proceed to discuss implications of our results.

\begin{deluxetable*}{lllcccrll} \label{tab:closest_WDs}
\tablenum{1}
\tablecaption{Closest 5 White Dwarfs with Comparable Surface Temperatures to WD 1856+534}
\tablewidth{0pt}
\tablehead{
WD identifier	&	T$_{\mbox{eff}}$			&	Age	&	Distance	&	log(L/L$_\odot$)	&	Radius	&	J	&	Spectral	&	Alias	\\
	&	(K)			&	(Gyr)	&	(pc)	&		&	R$_\odot$	&	mag	& Type	 	&	 	
}
\startdata \\[-8pt]
WD 0552-041	&	5182	$\pm$	81	&	7.31	&	6.441$^1$	&	-4.22	&	0.0097	&	13.05	&	DZ10.0	&	LHS 32	\\
WD 1334+039	&	4971	$\pm$	83	&	5.38	&	8.237$^2$	&	-4.03	&	0.0130	&	13.06	&	DA11	&	LHS 46	\\
WD 1132-325	&	5000	$\pm$	500	&	5.69	&	9.547$^1$	&	-4.05	&	0.0126	&	13.56	&	DC10	&	vB 4/LHS 309	\\
WD 0245+541	&	5139	$\pm$	82	&	6.51	&	10.866$^1$	&	-4.14	&	0.0107	&	13.87	&	DAZ9.5	&	LHS 1446	\\
WD 0821-669	&	5088	$\pm$	137	&	6.00	&	10.675$^1$	&	-4.10	&	0.0115	&	13.79	&	DA9.8	&	SCR J0821-6703	\\
\midrule
\textbf{WD 1856+534} &	\textbf{4710 $\pm$ 60} & \textbf{5.85} & \textbf{24.754$^1$} & \textbf{-4.13}	& \textbf{0.0131} &	\textbf{15.68} & \textbf{DC$^4$} & \textbf{LP 141-14}	\\
\enddata 
\tablecomments{WD 1856+534 is shown for comparison (bold entry). Ages and spectral types from \citet{Holberg2016}.}
\tablerefs{\cite{Gaia2018}$^1$, \citet{Holberg2008}$^{2}$, \citet{Holberg2016}$^3$, \citet{McCook2014}$^4$.}
\vspace{-10pt}
\end{deluxetable*}

\subsection{Atmospheric Characterization Potential for Closer White Dwarf Systems} 

Within 25\,pc of the Sun, there are more than 220 WDs, of which $\sim$ 24\% are multiple systems \citep{Holberg2016}. The closest, Sirius B (25,000\,K), resides 2.6\,pc away, with the closest single WD, van Maanen's Star (6,000\,K), at 4.3\,pc. Table~\ref{tab:closest_WDs} shows the closest 5 single WDs with similar surface temperatures to WD~1856+534, including WD~1856+534 for comparison. Due to their higher brightnesses (smaller magnitudes), planets orbiting closer WDs would present more promising characterization opportunities. For the closest WD in Table~\ref{tab:closest_WDs} (WD 0552-041), we estimate NIRSpec Prism error bars can be $\sim 5 \times$ smaller than for WD 1856+534. Therefore, our results for 25 transits of WD 1856+534 (i.e. strong detections of H$_2$O, CO$_2$, CH$_4$, and O$_3$) could be achieved with a \emph{single} transit of WD 0552-041.

\subsection{Comparing the M Dwarf and White Dwarf Opportunities}

Earth-like planets around cool small M dwarfs, such as TRAPPIST-1, are promising targets for characterization with the upcoming ELTs and JWST \citep[e.g.][]{Barstow2016,Lustig-Yaeger2019,Lin2020}. However, there remain outstanding challenges in interpreting transmission spectra of M dwarf terrestrial planets, notably contamination from unocculted starspots \citep[e.g.][]{Rackham2018}. Planets around WDs, on the other hand, are less susceptible to unocculted spots due to the greater WD area occulted by the planet during transit. WD planets also present increased planet to host radius ratios, consequently lowering the required time to achieve an SNR sufficient to remotely detect biosignatures, such as O$_3$ + CH$_4$ and O$_3$ + N$_2$O (e.g. Figure~\ref{fig:detections_vs_transits}). 

The relative potential of the M dwarf and WD opportunities are ultimately modulated by the number of characterizable targets. While the occurrence rate of rocky planets in the HZ of WDs is unknown, one can estimate the relative number of characterizable planets:

\begin{enumerate}
    \item \textbf{Number of stars / hosts}: \citet{Winters2019} report 48 M dwarfs within 5 pc, while \citet{Hollands2018} report 139 WDs within 20 pc. Comparing their respective local space-densities, WDs are $\sim 20 \times$ less common than M dwarfs.
    \item \textbf{Planetary occurrence rate}: M dwarf HZ terrestrial planet occurrence rate estimates range from $\approx$ 20-50 $\%$ \citep[e.g][]{Dressing2015,Hsu2020}. The equivalent rate for WDs is currently unknown, with K2 suggesting a limit of $< 28 \%$ \citep{vanSluijs2018}.
    \item \textbf{Transit probability}: Earth-sized planets in the M dwarf HZ have a $\sim$ 1-2$\%$ transit probability \citep[e.g.][]{Gillon2017,Dittmann2017}. The smaller $R_*$ of WDs and smaller HZ orbital separation yields similar transit probabilities of $\sim 1\%$. 
    \item \textbf{Characterization horizon}: detecting O$_3$, for example, would require $\gtrsim$ 100 JWST transits for a modern Earth atmosphere around the M dwarf TRAPPIST-1 (12.4\,pc) \citep{Lustig-Yaeger2019}. Our results for WD 1856+534 (24.8\,pc) show this can be achieved in 25 transits. Using PandExo, we estimated the limiting WD magnitude at which an O$_3$ detection would also require 100 transits. We found a J $\approx$ 17.7 ($\approx$ 62\,pc) WD is roughly equivalent to TRAPPIST-1e in atmospheric characterization potential. Earth-sized planets transiting WDs therefore offer a characterization horizon $\sim 5 \times$ further than for M dwarfs.  
\end{enumerate}

The greater characterization horizon for WD HZ planets overcomes the lower space-density of WD hosts. Our characterization horizon conservatively assumes a similar JWST observing time per transit for WD and M dwarf planets. Higher time efficiency per transit would expand this characterization volume. Assuming a comparable occurrence rate for temperate rocky planets in the WD HZ to the M dwarf HZ, the above factors suggest $\sim 6\times$ more characterizable transiting planets \correction{may} orbit in the WD HZ than the equivalent M dwarf HZ. \correction{Gaia DR2 has identified $\sim$ 4700 WDs within 62\,pc \citep{Gaia2018}, with some cool WDs likely missing from this sample. Assuming a 10\% occurrence rate, and 1\% transit probability, this would result in 5 WD systems with transiting Earth-sized planets within JWST's characterization horizon.}

\subsection{Characterizing White Dwarf Planet Atmospheres with Short Exposure Spectroscopy} \label{subsec:discussion_timing}

The transit duration of planets in the HZ of temperate WDs is remarkably short, lasting only several minutes for WD with surface temperatures below 6,000\,K \citep{Kozakis2018}. However, time constrained observations with JWST ($<$ 1\,hr start window) require at least a 1\,hr temporal baseline to correctly characterize instrument systematics. While this additional out-of-transit baseline can be used to access about 1/10$^{\rm th}$ of the full orbital lightcurve, it results in marginal improvements to the transmission spectrum precision. Present or future instruments capable of observing WD planet transits in short $\lesssim 10$\,min exposures would require $\sim 10 \times$ less observing time per transit compared to JWST. Ground-based facilities observing in atmospheric windows would provide one such avenue, capable of observing 100 transits with $\lesssim 20$\,hr exposure time. Short exposure spectroscopy of planets orbiting in the WD HZ provides a powerful avenue to rapidly conduct atmospheric reconnaissance of terrestrial exoplanets.

\subsection{Finding Rocky Planets in the White Dwarf Habitable Zone} \label{subsec:discussion_planet_detection}

The short transit duration of rocky planets in the HZ of cool ($\lesssim$ 6000\,K) WDs requires high cadence observations to enable their detection. Several ground-based and space-based surveys have undertaken preliminary searches \citep{Fulton2014,Xu2015,vanSluijs2018}. TESS's new 20\,s cadence mode will be instrumental to identify such planets in the near future. The large transit depths of these planets ($\gtrsim 50$\%) also renders them eminently detectable by upcoming high cadence ground-based surveys. \correction{For instance, the Vera Rubin Observatory (formerly LSST), even with relatively sparse sampling, could detect a subset of the transiting rocky planets in the WD HZ \citep{Cortes2019}.}

\subsection{The Search for Life in the Universe}

Our results demonstrate that terrestrial planets transiting white dwarfs offer an exceptional opportunity to characterize and remotely detect the presence of life on exoplanets. WD planets may be second-generation planets, forming after the stellar main sequence. Revealing their atmospheric composition will offer a treasure-trove of insights into planet formation, atmospheric chemistry, the lifespans of any biota, and the possibility of a second genesis after a star’s demise. 

While small planets around WDs have yet to be found, the detection of the first planetary-mass object around a WD \citep{Vanderburg2020} motivates the search for smaller planets around WDs with missions like TESS. \correction{With around 4700 WDs within 62\,pc \citep{Gaia2018}}, there are abundant opportunities to realize the white dwarf opportunity. Our search for life in the universe is bound to offer surprises. Somewhere, in the vast expanse of the cosmos, life may yet flourish; illuminated by the remnant core of a long-forgotten star.

\acknowledgments
\noindent We thank the referee for a fruitful dialogue and insightful suggestions, which improved the quality of this study. Support from the Carl Sagan Institute is gratefully acknowledged. Part of this research was carried out at the Jet Propulsion Laboratory, California Institute of Technology, under a contract with the National Aeronautics and Space Administration (NASA). This work was performed in part under contract with the California Institute of Technology/Jet Propulsion Laboratory, funded by NASA through the Sagan Fellowship Program executed by the NASA Exoplanet Science Institute.

\bibliography{WD_opportunity}{}

\begin{thebibliography}{}
\expandafter\ifx\csname natexlab\endcsname\relax\def\natexlab#1{#1}\fi
\providecommand{\url}[1]{\href{#1}{#1}}
\providecommand{\dodoi}[1]{doi:~\href{http://doi.org/#1}{\nolinkurl{#1}}}
\providecommand{\doeprint}[1]{\href{http://ascl.net/#1}{\nolinkurl{http://ascl.net/#1}}}
\providecommand{\doarXiv}[1]{\href{https://arxiv.org/abs/#1}{\nolinkurl{https://arxiv.org/abs/#1}}}

\bibitem[{{Agol}(2011)}]{Agol11}
{Agol}, E. 2011, \apjl, 731, L31, \dodoi{10.1088/2041-8205/731/2/L31}

\bibitem[{{Anglada-Escud{\'e}} {et~al.}(2016){Anglada-Escud{\'e}}, {Amado},
  {Barnes}, {Berdi{\~n}as}, {Butler}, {Coleman}, {de La Cueva}, {Dreizler},
  {Endl}, {Giesers}, {Jeffers}, {Jenkins}, {Jones}, {Kiraga}, {K{\"u}rster},
  {L{\'o}pez-Gonz{\'a}lez}, {Marvin}, {Morales}, {Morin}, {Nelson}, {Ortiz},
  {Ofir}, {Paardekooper}, {Reiners}, {Rodr{\'\i}guez},
  {Rodr{\'\i}guez-L{\'o}pez}, {Sarmiento}, {Strachan}, {Tsapras}, {Tuomi}, \&
  {Zechmeister}}]{Anglada2016}
{Anglada-Escud{\'e}}, G., {Amado}, P.~J., {Barnes}, J., {et~al.} 2016, Nature,
  536, 437, \dodoi{10.1038/nature19106}

\bibitem[{{Barstow} \& {Irwin}(2016)}]{Barstow2016}
{Barstow}, J.~K., \& {Irwin}, P.~G.~J. 2016, \mnras, 461, L92,
  \dodoi{10.1093/mnrasl/slw109}

\bibitem[{{Batalha} {et~al.}(2018){Batalha}, {Lewis}, {Line}, {Valenti}, \&
  {Stevenson}}]{Batalha2018}
{Batalha}, N.~E., {Lewis}, N.~K., {Line}, M.~R., {Valenti}, J., \& {Stevenson},
  K. 2018, \apjl, 856, L34, \dodoi{10.3847/2041-8213/aab896}

\bibitem[{{Batalha} {et~al.}(2017){Batalha}, {Mandell}, {Pontoppidan},
  {Stevenson}, {Lewis}, {Kalirai}, {Earl}, {Greene}, {Albert}, \&
  {Nielsen}}]{Batalha2017}
{Batalha}, N.~E., {Mandell}, A., {Pontoppidan}, K., {et~al.} 2017, \pasp, 129,
  064501, \dodoi{10.1088/1538-3873/aa65b0}

\bibitem[{{Benneke} \& {Seager}(2013)}]{Benneke2013}
{Benneke}, B., \& {Seager}, S. 2013, \apj, 778, 153,
  \dodoi{10.1088/0004-637X/778/2/153}

\bibitem[{{Benneke} {et~al.}(2019{\natexlab{a}}){Benneke}, {Knutson},
  {Lothringer}, {Crossfield}, {Moses}, {Morley}, {Kreidberg}, {Fulton},
  {Dragomir}, {Howard}, {Wong}, {D{\'e}sert}, {McCullough}, {Kempton},
  {Fortney}, {Gilliland }, {Deming}, \& {Kammer}}]{Benneke2019a}
{Benneke}, B., {Knutson}, H.~A., {Lothringer}, J., {et~al.} 2019{\natexlab{a}},
  Nature Astronomy, 3, 813, \dodoi{10.1038/s41550-019-0800-5}

\bibitem[{{Benneke} {et~al.}(2019{\natexlab{b}}){Benneke}, {Wong}, {Piaulet},
  {Knutson}, {Lothringer}, {Morley}, {Crossfield}, {Gao}, {Greene}, {Dressing},
  {Dragomir}, {Howard}, {McCullough}, {Kempton}, {Fortney}, \&
  {Fraine}}]{Benneke2019b}
{Benneke}, B., {Wong}, I., {Piaulet}, C., {et~al.} 2019{\natexlab{b}}, \apjl,
  887, L14, \dodoi{10.3847/2041-8213/ab59dc}

\bibitem[{{Bergeron} {et~al.}(2001){Bergeron}, {Leggett}, \& {Ruiz}}]{berg01}
{Bergeron}, P., {Leggett}, S.~K., \& {Ruiz}, M.~T. 2001, \apjs, 133, 413,
  \dodoi{10.1086/320356}

\bibitem[{{B{\'e}tr{\'e}mieux} \& {Kaltenegger}(2014)}]{Betremieux2014}
{B{\'e}tr{\'e}mieux}, Y., \& {Kaltenegger}, L. 2014, ApJ, 791, 7,
  \dodoi{10.1088/0004-637X/791/1/7}

\bibitem[{{Bonsor} {et~al.}(2020){Bonsor}, {Carter}, {Hollands},
  {G{\"a}nsicke}, {Leinhardt}, \& {Harrison}}]{Bonsor2020}
{Bonsor}, A., {Carter}, P.~J., {Hollands}, M., {et~al.} 2020, \mnras, 492,
  2683, \dodoi{10.1093/mnras/stz3603}

\bibitem[{{Buchner} {et~al.}(2014){Buchner}, {Georgakakis}, {Nandra}, {Hsu},
  {Rangel}, {Brightman}, {Merloni}, {Salvato}, {Donley}, \&
  {Kocevski}}]{Buchner2014}
{Buchner}, J., {Georgakakis}, A., {Nandra}, K., {et~al.} 2014, Astronomy and
  Astrophysics, 564, A125, \dodoi{10.1051/0004-6361/201322971}

\bibitem[{{Charbonneau} {et~al.}(2002){Charbonneau}, {Brown}, {Noyes}, \&
  {Gilliland}}]{Charbonneau2002}
{Charbonneau}, D., {Brown}, T.~M., {Noyes}, R.~W., \& {Gilliland}, R.~L. 2002,
  ApJ, 568, 377, \dodoi{10.1086/338770}

\bibitem[{{Cort{\'e}s} \& {Kipping}(2019)}]{Cortes2019}
{Cort{\'e}s}, J., \& {Kipping}, D. 2019, \mnras, 488, 1695,
  \dodoi{10.1093/mnras/stz1300}

\bibitem[{{Cutri} {et~al.}(2003){Cutri}, {Skrutskie}, {van Dyk}, {Beichman},
  {Carpenter}, {Chester}, {Cambresy}, {Evans}, {Fowler}, {Gizis}, {Howard},
  {Huchra}, {Jarrett}, {Kopan}, {Kirkpatrick}, {Light}, {Marsh}, {McCallon},
  {Schneider}, {Stiening}, {Sykes}, {Weinberg}, {Wheaton}, {Wheelock}, \&
  {Zacarias}}]{Cutri2003}
{Cutri}, R.~M., {Skrutskie}, M.~F., {van Dyk}, S., {et~al.} 2003, {2MASS All
  Sky Catalog of point sources.}

\bibitem[{{Debes} \& {Sigurdsson}(2002)}]{Debes2002}
{Debes}, J.~H., \& {Sigurdsson}, S. 2002, \apj, 572, 556,
  \dodoi{10.1086/340291}

\bibitem[{{Deming} {et~al.}(2013){Deming}, {Wilkins}, {McCullough}, {Burrows},
  {Fortney}, {Agol}, {Dobbs-Dixon}, {Madhusudhan}, {Crouzet}, {Desert},
  {Gilliland}, {Haynes}, {Knutson}, {Line}, {Magic}, {Mand ell}, {Ranjan},
  {Charbonneau}, {Clampin}, {Seager}, \& {Showman}}]{Deming2013}
{Deming}, D., {Wilkins}, A., {McCullough}, P., {et~al.} 2013, ApJ, 774, 95,
  \dodoi{10.1088/0004-637X/774/2/95}

\bibitem[{{Dittmann} {et~al.}(2017){Dittmann}, {Irwin}, {Charbonneau},
  {Bonfils}, {Astudillo-Defru}, {Haywood}, {Berta-Thompson}, {Newton},
  {Rodriguez}, {Winters}, {Tan}, {Almenara}, {Bouchy}, {Delfosse}, {Forveille},
  {Lovis}, {Murgas}, {Pepe}, {Santos}, {Udry}, {W{\"u}nsche}, {Esquerdo},
  {Latham}, \& {Dressing}}]{Dittmann2017}
{Dittmann}, J.~A., {Irwin}, J.~M., {Charbonneau}, D., {et~al.} 2017, \nat, 544,
  333, \dodoi{10.1038/nature22055}

\bibitem[{{Dressing} \& {Charbonneau}(2015)}]{Dressing2015}
{Dressing}, C.~D., \& {Charbonneau}, D. 2015, \apj, 807, 45,
  \dodoi{10.1088/0004-637X/807/1/45}

\bibitem[{{Feng} {et~al.}(2018){Feng}, {Robinson}, {Fortney}, {Lupu}, {Marley},
  {Lewis}, {Macintosh}, \& {Line}}]{Feng2018}
{Feng}, Y.~K., {Robinson}, T.~D., {Fortney}, J.~J., {et~al.} 2018, \aj, 155,
  200, \dodoi{10.3847/1538-3881/aab95c}

\bibitem[{{Feroz} \& {Hobson}(2008)}]{Feroz2008}
{Feroz}, F., \& {Hobson}, M.~P. 2008, MNRAS, 384, 449,
  \dodoi{10.1111/j.1365-2966.2007.12353.x}

\bibitem[{{Feroz} {et~al.}(2009){Feroz}, {Hobson}, \& {Bridges}}]{Feroz2009}
{Feroz}, F., {Hobson}, M.~P., \& {Bridges}, M. 2009, MNRAS, 398, 1601,
  \dodoi{10.1111/j.1365-2966.2009.14548.x}

\bibitem[{{Feroz} {et~al.}(2013){Feroz}, {Hobson}, {Cameron}, \&
  {Pettitt}}]{Feroz2013}
{Feroz}, F., {Hobson}, M.~P., {Cameron}, E., \& {Pettitt}, A.~N. 2013, arXiv
  e-prints, arXiv:1306.2144, \dodoi{10.21105/astro.1306.2144}

\bibitem[{{Fossati} {et~al.}(2012){Fossati}, {Bagnulo}, {Haswell}, {Patel},
  {Busuttil}, {Kowalski}, {Shulyak}, \& {Sterzik}}]{foss12}
{Fossati}, L., {Bagnulo}, S., {Haswell}, C.~A., {et~al.} 2012, \apjl, 757, L15,
  \dodoi{10.1088/2041-8205/757/1/L15}

\bibitem[{{Fulton} {et~al.}(2014){Fulton}, {Tonry}, {Flewelling}, {Burgett},
  {Chambers}, {Hodapp}, {Huber}, {Kaiser}, {Wainscoat}, \&
  {Waters}}]{Fulton2014}
{Fulton}, B.~J., {Tonry}, J.~L., {Flewelling}, H., {et~al.} 2014, \apj, 796,
  114, \dodoi{10.1088/0004-637X/796/2/114}

\bibitem[{{Gaia Collaboration} {et~al.}(2018){Gaia Collaboration}, {Brown},
  {Vallenari}, {Prusti}, {de Bruijne}, {Babusiaux}, {Bailer-Jones}, {Biermann},
  {Evans}, {Eyer}, {Jansen}, {Jordi}, {Klioner}, {Lammers}, {Lindegren},
  {Luri}, {Mignard}, {Panem}, {Pourbaix}, {Randich}, {Sartoretti}, {Siddiqui},
  {Soubiran}, {van Leeuwen}, {Walton}, {Arenou}, {Bastian}, {Cropper},
  {Drimmel}, {Katz}, {Lattanzi}, {Bakker}, {Cacciari}, {Casta{\~n}eda},
  {Chaoul}, {Cheek}, {De Angeli}, {Fabricius}, {Guerra}, {Holl}, {Masana},
  {Messineo}, {Mowlavi}, {Nienartowicz}, {Panuzzo}, {Portell}, {Riello},
  {Seabroke}, {Tanga}, {Th{\'e}venin}, {Gracia-Abril}, {Comoretto},
  {Garcia-Reinaldos}, {Teyssier}, {Altmann}, {Andrae}, {Audard},
  {Bellas-Velidis}, {Benson}, {Berthier}, {Blomme}, {Burgess}, {Busso},
  {Carry}, {Cellino}, {Clementini}, {Clotet}, {Creevey}, {Davidson}, {De
  Ridder}, {Delchambre}, {Dell'Oro}, {Ducourant},
  {Fern{\'a}ndez-Hern{\'a}ndez}, {Fouesneau}, {Fr{\'e}mat}, {Galluccio},
  {Garc{\'\i}a-Torres}, {Gonz{\'a}lez-N{\'u}{\~n}ez}, {Gonz{\'a}lez-Vidal},
  {Gosset}, {Guy}, {Halbwachs}, {Hambly}, {Harrison}, {Hern{\'a}ndez},
  {Hestroffer}, {Hodgkin}, {Hutton}, {Jasniewicz}, {Jean-Antoine-Piccolo},
  {Jordan}, {Korn}, {Krone-Martins}, {Lanzafame}, {Lebzelter}, {L{\"o}ffler},
  {Manteiga}, {Marrese}, {Mart{\'\i}n-Fleitas}, {Moitinho}, {Mora}, {Muinonen},
  {Osinde}, {Pancino}, {Pauwels}, {Petit}, {Recio-Blanco}, {Richards},
  {Rimoldini}, {Robin}, {Sarro}, {Siopis}, {Smith}, {Sozzetti}, {S{\"u}veges},
  {Torra}, {van Reeven}, {Abbas}, {Abreu Aramburu}, {Accart}, {Aerts},
  {Altavilla}, {{\'A}lvarez}, {Alvarez}, {Alves}, {Anderson}, {Andrei},
  {Anglada Varela}, {Antiche}, {Antoja}, {Arcay}, {Astraatmadja}, {Bach},
  {Baker}, {Balaguer-N{\'u}{\~n}ez}, {Balm}, {Barache}, {Barata}, {Barbato},
  {Barblan}, {Barklem}, {Barrado}, {Barros}, {Barstow}, {Bartholom{\'e}
  Mu{\~n}oz}, {Bassilana}, {Becciani}, {Bellazzini}, {Berihuete}, {Bertone},
  {Bianchi}, {Bienaym{\'e}}, {Blanco-Cuaresma}, {Boch}, {Boeche}, {Bombrun},
  {Borrachero}, {Bossini}, {Bouquillon}, {Bourda}, {Bragaglia}, {Bramante},
  {Breddels}, {Bressan}, {Brouillet}, {Br{\"u}semeister}, {Brugaletta},
  {Bucciarelli}, {Burlacu}, {Busonero}, {Butkevich}, {Buzzi}, {Caffau},
  {Cancelliere}, {Cannizzaro}, {Cantat-Gaudin}, {Carballo}, {Carlucci},
  {Carrasco}, {Casamiquela}, {Castellani}, {Castro-Ginard}, {Charlot},
  {Chemin}, {Chiavassa}, {Cocozza}, {Costigan}, {Cowell}, {Crifo}, {Crosta},
  {Crowley}, {Cuypers}, {Dafonte}, {Damerdji}, {Dapergolas}, {David}, {David},
  {de Laverny}, {De Luise}, {De March}, {de Martino}, {de Souza}, {de Torres},
  {Debosscher}, {del Pozo}, {Delbo}, {Delgado}, {Delgado}, {Di Matteo},
  {Diakite}, {Diener}, {Distefano}, {Dolding}, {Drazinos}, {Dur{\'a}n},
  {Edvardsson}, {Enke}, {Eriksson}, {Esquej}, {Eynard Bontemps}, {Fabre},
  {Fabrizio}, {Faigler}, {Falc{\~a}o}, {Farr{\`a}s Casas}, {Federici},
  {Fedorets}, {Fernique}, {Figueras}, {Filippi}, {Findeisen}, {Fonti},
  {Fraile}, {Fraser}, {Fr{\'e}zouls}, {Gai}, {Galleti}, {Garabato},
  {Garc{\'\i}a-Sedano}, {Garofalo}, {Garralda}, {Gavel}, {Gavras}, {Gerssen},
  {Geyer}, {Giacobbe}, {Gilmore}, {Girona}, {Giuffrida}, {Glass}, {Gomes},
  {Granvik}, {Gueguen}, {Guerrier}, {Guiraud}, {Guti{\'e}rrez-S{\'a}nchez},
  {Haigron}, {Hatzidimitriou}, {Hauser}, {Haywood}, {Heiter}, {Helmi}, {Heu},
  {Hilger}, {Hobbs}, {Hofmann}, {Holland}, {Huckle}, {Hypki}, {Icardi},
  {Jan{\ss}en}, {Jevardat de Fombelle}, {Jonker}, {Juh{\'a}sz}, {Julbe},
  {Karampelas}, {Kewley}, {Klar}, {Kochoska}, {Kohley}, {Kolenberg},
  {Kontizas}, {Kontizas}, {Koposov}, {Kordopatis}, {Kostrzewa-Rutkowska},
  {Koubsky}, {Lambert}, {Lanza}, {Lasne}, {Lavigne}, {Le Fustec}, {Le
  Poncin-Lafitte}, {Lebreton}, {Leccia}, {Leclerc}, {Lecoeur-Taibi},
  {Lenhardt}, {Leroux}, {Liao}, {Licata}, {Lindstr{\o}m}, {Lister}, {Livanou},
  {Lobel}, {L{\'o}pez}, {Managau}, {Mann}, {Mantelet}, {Marchal}, {Marchant},
  {Marconi}, {Marinoni}, {Marschalk{\'o}}, {Marshall}, {Martino}, {Marton},
  {Mary}, {Massari}, {Matijevi{\v{c}}}, {Mazeh}, {McMillan}, {Messina},
  {Michalik}, {Millar}, {Molina}, {Molinaro}, {Moln{\'a}r}, {Montegriffo},
  {Mor}, {Morbidelli}, {Morel}, {Morris}, {Mulone}, {Muraveva}, {Musella},
  {Nelemans}, {Nicastro}, {Noval}, {O'Mullane}, {Ord{\'e}novic},
  {Ord{\'o}{\~n}ez-Blanco}, {Osborne}, {Pagani}, {Pagano}, {Pailler},
  {Palacin}, {Palaversa}, {Panahi}, {Pawlak}, {Piersimoni}, {Pineau}, {Plachy},
  {Plum}, {Poggio}, {Poujoulet}, {Pr{\v{s}}a}, {Pulone}, {Racero}, {Ragaini},
  {Rambaux}, {Ramos-Lerate}, {Regibo}, {Reyl{\'e}}, {Riclet}, {Ripepi}, {Riva},
  {Rivard}, {Rixon}, {Roegiers}, {Roelens}, {Romero-G{\'o}mez}, {Rowell},
  {Royer}, {Ruiz-Dern}, {Sadowski}, {Sagrist{\`a} Sell{\'e}s}, {Sahlmann},
  {Salgado}, {Salguero}, {Sanna}, {Santana-Ros}, {Sarasso}, {Savietto},
  {Schultheis}, {Sciacca}, {Segol}, {Segovia}, {S{\'e}gransan}, {Shih},
  {Siltala}, {Silva}, {Smart}, {Smith}, {Solano}, {Solitro}, {Sordo}, {Soria
  Nieto}, {Souchay}, {Spagna}, {Spoto}, {Stampa}, {Steele},
  {Steidelm{\"u}ller}, {Stephenson}, {Stoev}, {Suess}, {Surdej}, {Szabados},
  {Szegedi-Elek}, {Tapiador}, {Taris}, {Tauran}, {Taylor}, {Teixeira},
  {Terrett}, {Teyssand ier}, {Thuillot}, {Titarenko}, {Torra Clotet}, {Turon},
  {Ulla}, {Utrilla}, {Uzzi}, {Vaillant}, {Valentini}, {Valette}, {van Elteren},
  {Van Hemelryck}, {van Leeuwen}, {Vaschetto}, {Vecchiato}, {Veljanoski},
  {Viala}, {Vicente}, {Vogt}, {von Essen}, {Voss}, {Votruba}, {Voutsinas},
  {Walmsley}, {Weiler}, {Wertz}, {Wevers}, {Wyrzykowski}, {Yoldas},
  {{\v{Z}}erjal}, {Ziaeepour}, {Zorec}, {Zschocke}, {Zucker}, {Zurbach}, \&
  {Zwitter}}]{Gaia2018}
{Gaia Collaboration}, {Brown}, A.~G.~A., {Vallenari}, A., {et~al.} 2018, \aap,
  616, A1, \dodoi{10.1051/0004-6361/201833051}

\bibitem[{{G{\"a}nsicke} {et~al.}(2019){G{\"a}nsicke}, {Schreiber}, {Toloza},
  {Fusillo}, {Koester}, \& {Manser}}]{gans19}
{G{\"a}nsicke}, B.~T., {Schreiber}, M.~R., {Toloza}, O., {et~al.} 2019, \nat,
  576, 61, \dodoi{10.1038/s41586-019-1789-8}

\bibitem[{{Gillon} {et~al.}(2017){Gillon}, {Triaud}, {Demory}, {Jehin}, {Agol},
  {Deck}, {Lederer}, {de Wit}, {Burdanov}, {Ingalls}, {Bolmont}, {Leconte},
  {Raymond}, {Selsis}, {Turbet}, {Barkaoui}, {Burgasser}, {Burleigh}, {Carey},
  {Chaushev}, {Copperwheat}, {Delrez}, {Fernandes}, {Holdsworth}, {Kotze}, {Van
  Grootel}, {Almleaky}, {Benkhaldoun}, {Magain}, \& {Queloz}}]{Gillon2017}
{Gillon}, M., {Triaud}, A. H.~M.~J., {Demory}, B.-O., {et~al.} 2017, Nature,
  542, 456, \dodoi{10.1038/nature21360}

\bibitem[{{Gordon} {et~al.}(2017){Gordon}, {Rothman}, {Hill}, {Kochanov},
  {Tan}, {Bernath}, {Birk}, {Boudon}, {Campargue}, {Chance}, {Drouin}, {Flaud},
  {Gamache}, {Hodges}, {Jacquemart}, {Perevalov}, {Perrin}, {Shine}, {Smith},
  {Tennyson}, {Toon}, {Tran}, {Tyuterev}, {Barbe}, {Cs{\'a}sz{\'a}r}, {Devi},
  {Furtenbacher}, {Harrison}, {Hartmann}, {Jolly}, {Johnson}, {Karman},
  {Kleiner}, {Kyuberis}, {Loos}, {Lyulin}, {Massie}, {Mikhailenko},
  {Moazzen-Ahmadi}, {M{\"u}ller}, {Naumenko}, {Nikitin}, {Polyansky}, {Rey},
  {Rotger}, {Sharpe}, {Sung}, {Starikova}, {Tashkun}, {Auwera}, {Wagner},
  {Wilzewski}, {Wcis{\l}o}, {Yu}, \& {Zak}}]{Gordon2017}
{Gordon}, I.~E., {Rothman}, L.~S., {Hill}, C., {et~al.} 2017, \jqsrt, 203, 3,
  \dodoi{10.1016/j.jqsrt.2017.06.038}

\bibitem[{{Hoeijmakers} {et~al.}(2018){Hoeijmakers}, {Ehrenreich}, {Heng},
  {Kitzmann}, {Grimm}, {Allart}, {Deitrick}, {Wyttenbach}, {Oreshenko}, {Pino},
  {Rimmer}, {Molinari}, \& {Di Fabrizio}}]{Hoeijmakers2018}
{Hoeijmakers}, H.~J., {Ehrenreich}, D., {Heng}, K., {et~al.} 2018, Nature, 560,
  453, \dodoi{10.1038/s41586-018-0401-y}

\bibitem[{{Holberg} {et~al.}(2008){Holberg}, {Bergeron}, \&
  {Gianninas}}]{Holberg2008}
{Holberg}, J.~B., {Bergeron}, P., \& {Gianninas}, A. 2008, \aj, 135, 1239,
  \dodoi{10.1088/0004-6256/135/4/1239}

\bibitem[{{Holberg} {et~al.}(2016){Holberg}, {Oswalt}, {Sion}, \&
  {McCook}}]{Holberg2016}
{Holberg}, J.~B., {Oswalt}, T.~D., {Sion}, E.~M., \& {McCook}, G.~P. 2016,
  \mnras, 462, 2295, \dodoi{10.1093/mnras/stw1357}

\bibitem[{{Hollands} {et~al.}(2018){Hollands}, {Tremblay}, {G{\"a}nsicke},
  {Gentile-Fusillo}, \& {Toonen}}]{Hollands2018}
{Hollands}, M.~A., {Tremblay}, P.~E., {G{\"a}nsicke}, B.~T., {Gentile-Fusillo},
  N.~P., \& {Toonen}, S. 2018, \mnras, 480, 3942, \dodoi{10.1093/mnras/sty2057}

\bibitem[{{Howell} {et~al.}(2014){Howell}, {Sobeck}, {Haas}, {Still},
  {Barclay}, {Mullally}, {Troeltzsch}, {Aigrain}, {Bryson}, {Caldwell},
  {Chaplin}, {Cochran}, {Huber}, {Marcy}, {Miglio}, {Najita}, {Smith},
  {Twicken}, \& {Fortney}}]{Howell2014}
{Howell}, S.~B., {Sobeck}, C., {Haas}, M., {et~al.} 2014, \pasp, 126, 398,
  \dodoi{10.1086/676406}

\bibitem[{{Hsu} {et~al.}(2020){Hsu}, {Ford}, \& {Terrien}}]{Hsu2020}
{Hsu}, D.~C., {Ford}, E.~B., \& {Terrien}, R. 2020, arXiv e-prints,
  arXiv:2002.02573.
\newblock \doarXiv{2002.02573}

\bibitem[{{Irwin} {et~al.}(2018){Irwin}, {Toledo}, {Garland}, {Teanby},
  {Fletcher}, {Orton}, \& {B{\'e}zard}}]{Irwin2018}
{Irwin}, P. G.~J., {Toledo}, D., {Garland}, R., {et~al.} 2018, Nature
  Astronomy, 2, 420, \dodoi{10.1038/s41550-018-0432-1}

\bibitem[{{Jura} \& {Young}(2014)}]{Jura2014}
{Jura}, M., \& {Young}, E.~D. 2014, Annual Review of Earth and Planetary
  Sciences, 42, 45, \dodoi{10.1146/annurev-earth-060313-054740}

\bibitem[{{Kaltenegger}(2017)}]{kalt17}
{Kaltenegger}, L. 2017, \araa, 55, 433,
  \dodoi{10.1146/annurev-astro-082214-122238}

\bibitem[{{Kaltenegger} \& {Sasselov}(2010)}]{Kaltenegger2010}
{Kaltenegger}, L., \& {Sasselov}, D. 2010, \apj, 708, 1162,
  \dodoi{10.1088/0004-637X/708/2/1162}

\bibitem[{{Karman} {et~al.}(2019){Karman}, {Gordon}, {van der Avoird},
  {Baranov}, {Boulet}, {Drouin}, {Groenenboom}, {Gustafsson}, {Hartmann},
  {Kurucz}, {Rothman}, {Sun}, {Sung}, {Thalman}, {Tran}, {Wishnow},
  {Wordsworth}, {Vigasin}, {Volkamer}, \& {van der Zande}}]{Karman2019}
{Karman}, T., {Gordon}, I.~E., {van der Avoird}, A., {et~al.} 2019, \icarus,
  328, 160, \dodoi{10.1016/j.icarus.2019.02.034}

\bibitem[{{Kepler} {et~al.}(2016){Kepler}, {Pelisoli}, {Koester}, {Ourique},
  {Romero}, {Reindl}, {Kleinman}, {Eisenstein}, {Valois}, \&
  {Amaral}}]{Kepler2016}
{Kepler}, S.~O., {Pelisoli}, I., {Koester}, D., {et~al.} 2016, \mnras, 455,
  3413, \dodoi{10.1093/mnras/stv2526}

\bibitem[{{Kilpatrick} {et~al.}(2018){Kilpatrick}, {Cubillos}, {Stevenson},
  {Lewis}, {Wakeford}, {MacDonald}, {Madhusudhan}, {Blecic}, {Bruno},
  {Burrows}, {Deming}, {Heng}, {Line}, {Morley}, {Parmentier}, {Tucker},
  {Valenti}, {Waldmann}, {Bean}, {Beichman}, {Fraine}, {Krick}, {Lothringer},
  \& {Mandell}}]{Kilpatrick2018}
{Kilpatrick}, B.~M., {Cubillos}, P.~E., {Stevenson}, K.~B., {et~al.} 2018, \aj,
  156, 103, \dodoi{10.3847/1538-3881/aacea7}

\bibitem[{{Kochanov} {et~al.}(2016){Kochanov}, {Gordon}, {Rothman},
  {Wcis{\l}o}, {Hill}, \& {Wilzewski}}]{Kochanov2016}
{Kochanov}, R.~V., {Gordon}, I.~E., {Rothman}, L.~S., {et~al.} 2016, \jqsrt,
  177, 15, \dodoi{10.1016/j.jqsrt.2016.03.005}

\bibitem[{{Koester} {et~al.}(2014){Koester}, {G{\"a}nsicke}, \&
  {Farihi}}]{Koester2014}
{Koester}, D., {G{\"a}nsicke}, B.~T., \& {Farihi}, J. 2014, \aap, 566, A34,
  \dodoi{10.1051/0004-6361/201423691}

\bibitem[{{Kozakis} {et~al.}(2018){Kozakis}, {Kaltenegger}, \&
  {Hoard}}]{Kozakis2018}
{Kozakis}, T., {Kaltenegger}, L., \& {Hoard}, D.~W. 2018, \apj, 862, 69,
  \dodoi{10.3847/1538-4357/aacbc7}

\bibitem[{{Kozakis} {et~al.}(2020){Kozakis}, {Lin}, \&
  {Kaltenegger}}]{Kozakis2020}
{Kozakis}, T., {Lin}, Z., \& {Kaltenegger}, L. 2020, \apjl, 894, L6,
  \dodoi{10.3847/2041-8213/ab6f6a}

\bibitem[{{Kreidberg} {et~al.}(2019){Kreidberg}, {Koll}, {Morley}, {Hu},
  {Schaefer}, {Deming}, {Stevenson}, {Dittmann}, {Vanderburg}, {Berardo},
  {Guo}, {Stassun}, {Crossfield}, {Charbonneau}, {Latham}, {Loeb}, {Ricker},
  {Seager}, \& {Vand erspek}}]{Kreidberg2019}
{Kreidberg}, L., {Koll}, D. D.~B., {Morley}, C., {et~al.} 2019, \nat, 573, 87,
  \dodoi{10.1038/s41586-019-1497-4}

\bibitem[{{Krissansen-Totton} {et~al.}(2018){Krissansen-Totton}, {Garland},
  {Irwin}, \& {Catling}}]{Krissansen-Totton2018}
{Krissansen-Totton}, J., {Garland}, R., {Irwin}, P., \& {Catling}, D.~C. 2018,
  \aj, 156, 114, \dodoi{10.3847/1538-3881/aad564}

\bibitem[{{Lin} \& {Kaltenegger}(2020)}]{Lin2020}
{Lin}, Z., \& {Kaltenegger}, L. 2020, \mnras, 491, 2845,
  \dodoi{10.1093/mnras/stz3213}

\bibitem[{{Loeb} \& {Maoz}(2013)}]{loeb13}
{Loeb}, A., \& {Maoz}, D. 2013, \mnras, 432, L11, \dodoi{10.1093/mnrasl/slt026}

\bibitem[{{Lustig-Yaeger} {et~al.}(2019){Lustig-Yaeger}, {Meadows}, \&
  {Lincowski}}]{Lustig-Yaeger2019}
{Lustig-Yaeger}, J., {Meadows}, V.~S., \& {Lincowski}, A.~P. 2019, \aj, 158,
  27, \dodoi{10.3847/1538-3881/ab21e0}

\bibitem[{{Macdonald} \& {Cowan}(2019)}]{MacdonaldE2019}
{Macdonald}, E. J.~R., \& {Cowan}, N.~B. 2019, \mnras, 489, 196,
  \dodoi{10.1093/mnras/stz2047}

\bibitem[{{MacDonald} \& {Madhusudhan}(2017)}]{MacDonald2017}
{MacDonald}, R.~J., \& {Madhusudhan}, N. 2017, \mnras, 469, 1979,
  \dodoi{10.1093/mnras/stx804}

\bibitem[{{MacDonald} \& {Madhusudhan}(2019)}]{MacDonald2019}
---. 2019, \mnras, 486, 1292, \dodoi{10.1093/mnras/stz789}

\bibitem[{{Madhusudhan}(2018)}]{Madhusudhan2018}
{Madhusudhan}, N. 2018, {Atmospheric Retrieval of Exoplanets} (Springer
  International Publishing AG), 104, \dodoi{10.1007/978-3-319-55333-7_104}

\bibitem[{{Madhusudhan} \& {Seager}(2009)}]{Madhusudhan2009}
{Madhusudhan}, N., \& {Seager}, S. 2009, ApJ, 707, 24,
  \dodoi{10.1088/0004-637X/707/1/24}

\bibitem[{{McCook} \& {Sion}(2016)}]{McCook2014}
{McCook}, G.~P., \& {Sion}, E.~M. 2016, VizieR Online Data Catalog, B/wd

\bibitem[{{McCree}(1971)}]{mccr71}
{McCree}, K.~J. 1971, Agricultural Meterology, 9, 191

\bibitem[{{Meadows} {et~al.}(2018){Meadows}, {Arney}, {Schwieterman},
  {Lustig-Yaeger}, {Lincowski}, {Robinson}, {Domagal-Goldman}, {Deitrick},
  {Barnes}, {Fleming}, {Luger}, {Driscoll}, {Quinn}, \& {Crisp}}]{Meadows2018}
{Meadows}, V.~S., {Arney}, G.~N., {Schwieterman}, E.~W., {et~al.} 2018,
  Astrobiology, 18, 133, \dodoi{10.1089/ast.2016.1589}

\bibitem[{{Morley} {et~al.}(2017){Morley}, {Kreidberg}, {Rustamkulov},
  {Robinson}, \& {Fortney}}]{Morley2017}
{Morley}, C.~V., {Kreidberg}, L., {Rustamkulov}, Z., {Robinson}, T., \&
  {Fortney}, J.~J. 2017, \apj, 850, 121, \dodoi{10.3847/1538-4357/aa927b}

\bibitem[{{O'Malley-James} {et~al.}(2013){O'Malley-James}, {Greaves}, {Raven},
  \& {Cockell}}]{omal2013}
{O'Malley-James}, J.~T., {Greaves}, J.~S., {Raven}, J.~A., \& {Cockell}, C.~S.
  2013, International Journal of Astrobiology, 12, 99,
  \dodoi{10.1017/S147355041200047X}

\bibitem[{{Rackham} {et~al.}(2018){Rackham}, {Apai}, \&
  {Giampapa}}]{Rackham2018}
{Rackham}, B.~V., {Apai}, D., \& {Giampapa}, M.~S. 2018, \apj, 853, 122,
  \dodoi{10.3847/1538-4357/aaa08c}

\bibitem[{{Robinson} {et~al.}(2017){Robinson}, {Fortney}, \&
  {Hubbard}}]{Robinson2017b}
{Robinson}, T.~D., {Fortney}, J.~J., \& {Hubbard}, W.~B. 2017, ApJ, 850, 128,
  \dodoi{10.3847/1538-4357/aa951e}

\bibitem[{{Sagan} {et~al.}(1993){Sagan}, {Thompson}, {Carlson}, {Gurnett}, \&
  {Hord}}]{Sagan1993}
{Sagan}, C., {Thompson}, W.~R., {Carlson}, R., {Gurnett}, D., \& {Hord}, C.
  1993, \nat, 365, 715, \dodoi{10.1038/365715a0}

\bibitem[{{Saumon} {et~al.}(2014){Saumon}, {Holberg}, \&
  {Kowalski}}]{Saumon2014}
{Saumon}, D., {Holberg}, J.~B., \& {Kowalski}, P.~M. 2014, \apj, 790, 50,
  \dodoi{10.1088/0004-637X/790/1/50}

\bibitem[{{Schwieterman} {et~al.}(2015){Schwieterman}, {Robinson}, {Meadows},
  {Misra}, \& {Domagal-Goldman}}]{Schwieterman2015}
{Schwieterman}, E.~W., {Robinson}, T.~D., {Meadows}, V.~S., {Misra}, A., \&
  {Domagal-Goldman}, S. 2015, \apj, 810, 57, \dodoi{10.1088/0004-637X/810/1/57}

\bibitem[{{Sedaghati} {et~al.}(2017){Sedaghati}, {Boffin}, {MacDonald},
  {Gandhi}, {Madhusudhan}, {Gibson}, {Oshagh}, {Claret}, \&
  {Rauer}}]{Sedaghati2017}
{Sedaghati}, E., {Boffin}, H. M.~J., {MacDonald}, R.~J., {et~al.} 2017, Nature,
  549, 238, \dodoi{10.1038/nature23651}

\bibitem[{{Serdyuchenko} {et~al.}(2014){Serdyuchenko}, {Gorshelev}, {Weber},
  {Chehade}, \& {Burrows}}]{Serdyuchenko2014}
{Serdyuchenko}, A., {Gorshelev}, V., {Weber}, M., {Chehade}, W., \& {Burrows},
  J.~P. 2014, Atmospheric Measurement Techniques, 7, 625,
  \dodoi{10.5194/amt-7-625-2014}

\bibitem[{{Tremblay} {et~al.}(2020){Tremblay}, {Line}, {Stevenson}, {Kataria},
  {Zellem}, {Fortney}, \& {Morley}}]{Tremblay2020}
{Tremblay}, L., {Line}, M.~R., {Stevenson}, K., {et~al.} 2020, \aj, 159, 117,
  \dodoi{10.3847/1538-3881/ab64dd}

\bibitem[{{Trotta}(2017)}]{Trotta2017}
{Trotta}, R. 2017, arXiv e-prints, arXiv:1701.01467.
\newblock \doarXiv{1701.01467}

\bibitem[{{Tsiaras} {et~al.}(2019){Tsiaras}, {Waldmann}, {Tinetti}, {Tennyson},
  \& {Yurchenko}}]{Tsiaras2019}
{Tsiaras}, A., {Waldmann}, I.~P., {Tinetti}, G., {Tennyson}, J., \&
  {Yurchenko}, S.~N. 2019, Nature Astronomy, 3, 1086,
  \dodoi{10.1038/s41550-019-0878-9}

\bibitem[{{van Sluijs} \& {Van Eylen}(2018)}]{vanSluijs2018}
{van Sluijs}, L., \& {Van Eylen}, V. 2018, \mnras, 474, 4603,
  \dodoi{10.1093/mnras/stx3068}

\bibitem[{{Vanderburg} {et~al.}(2020){Vanderburg}, {Rappaport}, {Xu},
  {Crossfield}, {Becker}, {Gary}, {Murgas}, \& {et}}]{Vanderburg2020}
{Vanderburg}, A., {Rappaport}, S.~A., {Xu}, S., {et~al.} 2020, Nature,
  \dodoi{10.1038/s41586-020-2713-y}

\bibitem[{{Vanderburg} {et~al.}(2015){Vanderburg}, {Johnson}, {Rappaport},
  {Bieryla}, {Irwin}, {Lewis}, {Kipping}, {Brown}, {Dufour}, {Ciardi}, {Angus},
  {Schaefer}, {Latham}, {Charbonneau}, {Beichman}, {Eastman}, {McCrady},
  {Wittenmyer}, \& {Wright}}]{Vanderburg2015}
{Vanderburg}, A., {Johnson}, J.~A., {Rappaport}, S., {et~al.} 2015, \nat, 526,
  546, \dodoi{10.1038/nature15527}

\bibitem[{{Veras} \& {Fuller}(2019)}]{Veras2019a}
{Veras}, D., \& {Fuller}, J. 2019, \mnras, 489, 2941,
  \dodoi{10.1093/mnras/stz2339}

\bibitem[{{Veras} \& {G{\"a}nsicke}(2015)}]{Veras2015}
{Veras}, D., \& {G{\"a}nsicke}, B.~T. 2015, \mnras, 447, 1049,
  \dodoi{10.1093/mnras/stu2475}

\bibitem[{{Veras} {et~al.}(2019){Veras}, {Efroimsky}, {Makarov}, {Bou{\'e}},
  {Wolthoff}, {Reffert}, {Quirrenbach}, {Tremblay}, \&
  {G{\"a}nsicke}}]{Veras2019b}
{Veras}, D., {Efroimsky}, M., {Makarov}, V.~V., {et~al.} 2019, \mnras, 486,
  3831, \dodoi{10.1093/mnras/stz965}

\bibitem[{{Winters} {et~al.}(2019){Winters}, {Henry}, {Jao}, {Subasavage},
  {Chatelain}, {Slatten}, {Riedel}, {Silverstein}, \& {Payne}}]{Winters2019}
{Winters}, J.~G., {Henry}, T.~J., {Jao}, W.-C., {et~al.} 2019, \aj, 157, 216,
  \dodoi{10.3847/1538-3881/ab05dc}

\bibitem[{{Xu} {et~al.}(2015){Xu}, {Ertel}, {Wahhaj}, {Milli}, {Scicluna}, \&
  {Bertrang}}]{Xu2015}
{Xu}, S., {Ertel}, S., {Wahhaj}, Z., {et~al.} 2015, \aap, 579, L8,
  \dodoi{10.1051/0004-6361/201526179}

\end{thebibliography}

\bibliographystyle{aasjournal}

\end{document}